\documentclass[amsmath,amssymb, aps, prx, longbibliography, twocolumn]{revtex4-1}

\usepackage{natbib}
\usepackage{graphicx}
\usepackage{dcolumn}
\usepackage{bm}
\usepackage[usenames]{color}

\begin{document}

\title{ Probing transport in quantum many-fermion simulations\\ via quantum loop topography}

\author{Yi Zhang$^1$}
\email{frankzhangyi@gmail.com}
\author{Carsten Bauer$^2$}
\author{Peter Broecker$^2$}
\author{Simon Trebst$^2$}
\author{Eun-Ah Kim$^1$}
\email{eun-ah.kim@cornell.edu}

\affiliation{%
$^{1}$Department of Physics, Cornell University, Ithaca, New York 14853, USA}%
\affiliation{$^{2}$Institute for Theoretical Physics, University of Cologne, 50937 Cologne, Germany}

\date{\today}

\begin{abstract}
Quantum many-fermion systems give rise to diverse states of matter that often reveal themselves in distinctive transport properties. While some of these states can be captured by microscopic models accessible to numerical exact quantum Monte Carlo simulations, it nevertheless remains challenging to  numerically access their transport properties.
Here we demonstrate that quantum loop topography (QLT) can be used to directly probe transport by machine learning current-current correlations in imaginary time.
We showcase this approach by studying the emergence of superconducting fluctuations in the negative-U Hubbard model and
a spin-fermion model for a metallic quantum critical point.
For both sign-free models, we find that the QLT approach detects a change in transport in very good agreement with their established
phase diagrams.
These proof-of-principle calculations combined with the numerical efficiency of the QLT approach point a way to identify hitherto elusive
transport phenomena such as non-Fermi liquids using machine learning algorithms.
\end{abstract}

\maketitle


Quantum many-body systems exhibit an intriguing diversity of collective states that have no classical counterpart.
Paradigmatic examples include the formation of Bose-Einstein condensates and superfluids in bosonic systems \cite{BEC},
the emergence of spin liquids and macroscopic entanglement in magnetic systems \cite{Balents2010},
or the observation of superconductivity in many-electron systems \cite{QuantumLiquids}.
The microscopic physics giving rise to these phenomena is well understood for systems
of many interacting bosonic or spin degrees of freedom,
either via controlled analytical calculations for minimal model Hamiltonians or via numerical simulations
providing even quantitative guidance. The fundamental understanding of quantum many-fermion systems, however,
has proved to be more elusive. The distinct feature leading to a seemingly unsurmountable complication
for these systems arguably is the profusion of gapless modes near the Fermi energy.
On the analytical side, the concurrent treatment of these gapless degrees of freedom and their interactions with other (bosonic) soft modes,
e.g.~in the vicinity of a quantum phase transition \cite{Hertz1976, Millis1993}, has remained a formidable challenge.
On the numerical side, many-fermion systems have long proved to resist a solution via quantum Monte Carlo techniques
due to the occurrence of the so-called sign problem \cite{Loh1990} that is closely linked to the complex sign structure of the
many-fermion wavefunction (another consequence of the existence of a multitude of gapless modes).
Adding to this complexity, the key features revealing the nature of collective many-fermion states
(such as superconductors, strange metals, or non-Fermi liquids) are often their transport properties
that are notoriously difficult to calculate.

It is the purpose of this manuscript to outline a numerical scheme that
allows for a direct quantitative probe of transport properties in interacting many-electron systems
by combining elements from machine learning and quantum Monte Carlo (QMC)
techniques.
To do so, we build on progress on two separate fronts advancing the numerical description of many-fermion systems.
First, it has been realized that quantum criticality in itinerant fermion systems can be studied in a numerically exact manner
in sign-problem free models
\cite{Berg2018}
built around the effective action for multiple fermion bands.
However, to infer transport properties one faces the problem that QMC simulations intrinsically provide access to imaginary time correlations only,
and the analytic continuation to real time is numerically ill-posed, yielding no controlled framework
to probe transport properties.
Instead, we resort to the recent development of machine learning approaches in quantum statistical physics and demonstrate that quantum loop topography (QLT), a numerical scheme initially designed to identify the topological Hall response of a system~\cite{qlt2016}, can in fact be used to measure {\em longitudinal} transport properties of itinerant many-fermion systems.

Here we show that the QLT approach can be adapted to extract the essential features of the imaginary time current-current correlation function. 
One principle example which we focus on is the study of superconductivity, whose onset can, in principle, be tracked e.g. via the superfluid density which can be rigorously obtained from current-current correlations \cite{Scalapino1992, Scalapino1993}.
We demonstrate that the QLT+QMC approach succeeds in identifying the essential features of this transition without any prior knowledge (e.g. about the explicit calculation of the superfluid density)
and quantitatively matches existing results for the onset of superconductivity for a number of microscopic model systems,
but at a considerably lower computational cost.

\begin{figure}[t]
    \centering
    \includegraphics[scale=.85]{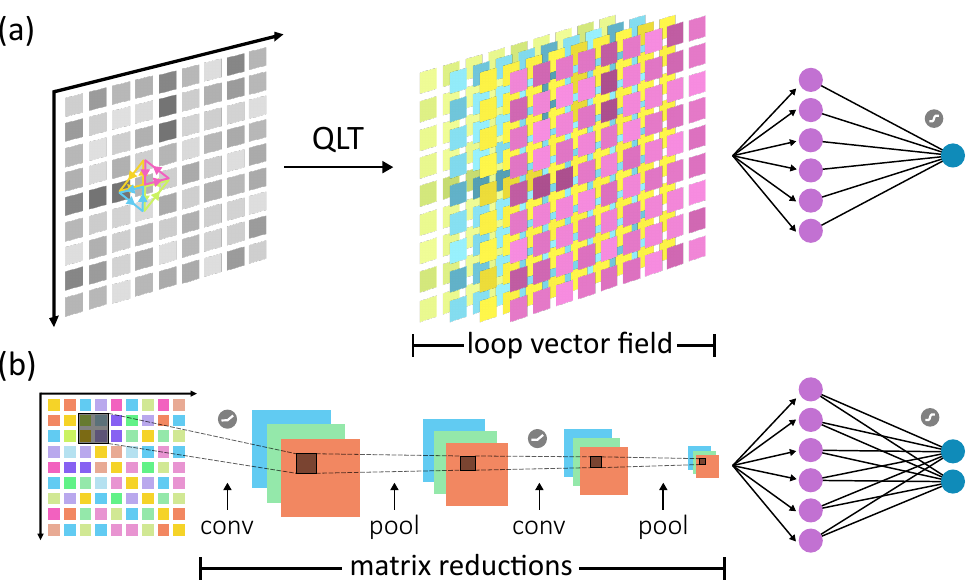}
    \caption{{\bf Neural network architectures}.
    			(a) QLT used as an input to a feed-forward fully-connected shallow neural network with one hidden layer
			consisting of $n=40$ sigmoid neurons. Only triangular loops $ L^\triangle_{jkl}$ are illustrated.
    			(b) Deep convolutional neural network that convolves and pools the unprocessed Green's functions
				$P(\mathbf{r}, \mathbf{r}')$ before threading them through a fully-connected layer of $n=256$ hidden neurons.}
    \label{fig:network_architecture}
\end{figure}

\noindent{\em QLT for longitudinal transport.-- }
The recent foray of applying machine learning techniques to quantum many-body systems can roughly be divided into two classes of general approaches:
(i) the representation of many-body wavefunctions using restricted Boltzmann machines allowing for a new class of variational algorithms to efficiently find ground states of quantum many-body systems
\cite{Carleo2016, Deng2016, Deng2017, Glasser2018, Torlai2018, Gaoxun2017}, and
(ii) the use of artificial neural networks (ANNs), typically combined with preprocessing steps, to allow for quantum state recognition
\cite{Melko20161, Nieuwenburg2017, LeiWang2016, Simon2016, Kelvin2016, Ohtsuki2016, Ohtsuki2017, Titus2017, qlt2016, FrankMLZ2, Broecker2017, ZhaiHui2017, Iakovlev2018, PollmannML2018, Anna2018}.
In the latter category, the QLT~\cite{qlt2016} stands out as a preprocessing step that, by using loop topography as a filter,  selects and organizes the simulation data with the {\em physical response} characteristic of the target phase in mind (and thereby distinguishes itself from e.g.~the application of convolutional neural networks (CNNs) whose motivation is primarily rooted in image recognition techniques).
The QLT-preprocessed data is then fed into a shallow ANN, which can be trained to discriminate different quantum phases of matter. This general setup is schematically illustrated in Fig.~\ref{fig:network_architecture}.
The QLT approach has so far been employed to the detection of topological order in integer and fractional Chern insulators~\cite{qlt2016} by targeting the Hall transport and to positively identify a $\mathbb{Z}_2$ spin liquid~\cite{Zhang2017} by targeting Wilson loops.

\begin{figure}[b]
\includegraphics[scale=0.48]{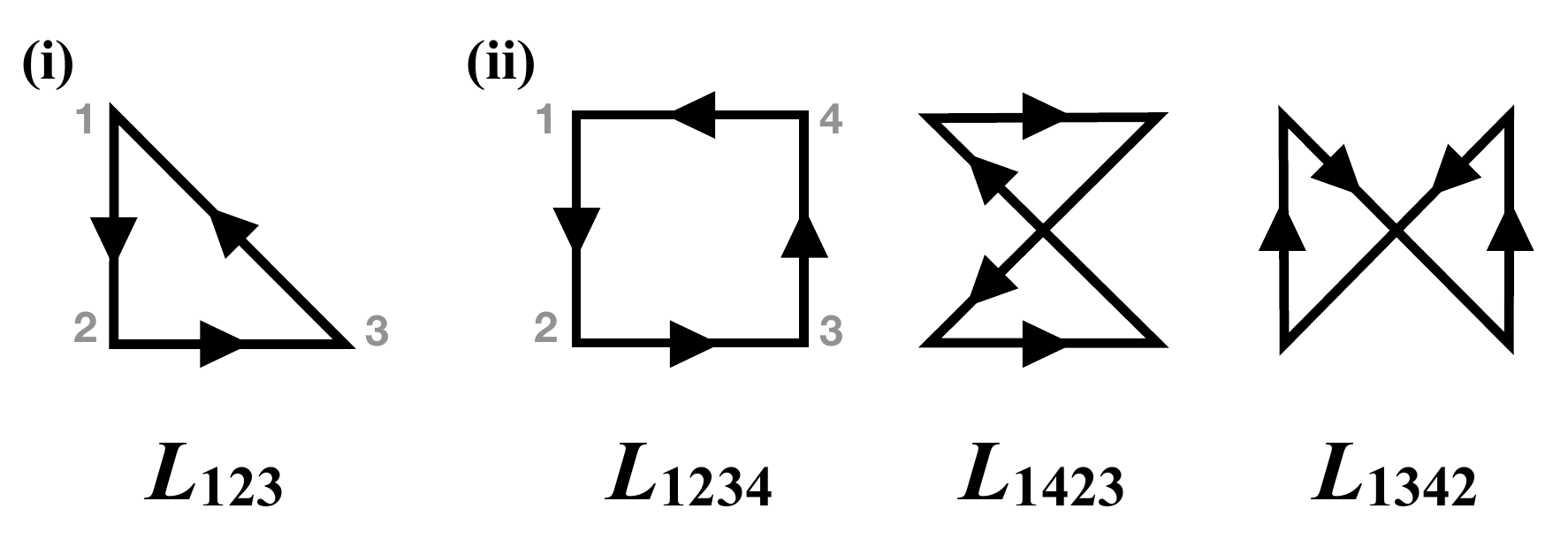}
\caption{
 Illustration of the (i) triangular and (ii) quadrilateral {\bf loop operators} employed to calculate the longitudinal transport.
}\label{fig:loops}
\end{figure}

Targeting the longitudinal transport for the purpose of the current study, we build a vector at each site $j$ consisting of all small loops with three vertices, $L^\triangle_{jkl}$ and with four vertices, $L^\Box_{jklm}$ including the site $j$. The loops represent
chained products of Green's functions, i.e. bilinear fermionic operators  $c_i^\dagger c_j^{\phantom\dagger}$, evaluated for a given Monte Carlo sample
$\alpha$, $\widetilde{P}_{jk}|_{\alpha}$:
\begin{equation}
    L^\triangle_{jkl}\equiv\widetilde{P}_{jk}|_{\alpha} \widetilde{P}_{kl}|_{\beta} \widetilde{P}_{lj}|_{\gamma} \,,
    \label{eq:triangle}
\end{equation}
and
\begin{equation}
L^\Box_{jklm}\equiv\widetilde{P}_{jk}|_{\alpha'} \widetilde{P}_{kl}|_{\beta'} \widetilde{P}_{lm}|_{\gamma'} \widetilde{P}_{mj}|_{\delta'} \,,
\label{eq:quad}
\end{equation}
limiting the neighboring sites to be within a short-distance cutoff $d_c$.
The loop operators associated with a site are illustrated in Fig.~\ref{fig:loops} for the shortest lengths, i.e. length 3 and 4.

To see how the loop operators $L^\Box_{jklm}$ and $ L^\triangle_{jkl}$ capture the longitudinal transport, consider the zero-frequency current-current correlation function
\begin{equation}
\Lambda_{xx}({\bf r}_1,{\bf r}_2;\omega_n=0)\equiv\int d\tau\left\langle \hat{j}_x\left({\bf r}_1, \tau \right) \hat{j}_x\left({\bf r}_2, 0 \right)\right\rangle,
\label{eq:Lambda}
\end{equation}
where $\hat{j}_x({\bf r}_1,\tau)=e^{H\tau} \hat{j}_x({\bf r}_1) e^{-H\tau}$ with the current density operator $\hat{j}_x({\bf r_1})= -i[H({\bf r}_1),\hat{x}]$.
Its Fourier transform 
is well known to be related \cite{Scalapino1993, Scalapino1992} to the superfluid density $\rho_s$ through $\rho_s\propto \Lambda_{xx}(q_x\!\rightarrow\! 0,q_y\!=\!0,\omega_n\!=\!0)-\Lambda_{xx}(q_x\!=\!0,q_y\!\rightarrow\!0,\omega_n\!=\!0)$.

To gain further analytical insight, consider
a gapped mean-field Hamiltonian with a single flat band which can be approximated as $H'=-\Pi$, where
$\Pi\equiv |G\rangle \langle G|$ is the projection operator for the ground state $|G\rangle$. At zero temperature we can evaluate the current-current correlation function for the system with the Hamiltonian ${H}'$:
\begin{equation}
\left. \begin{aligned}
\Lambda_{xx}(&{\bf r}_1,{\bf r}_2;\omega_n\!=\!0)
= \langle G|\hat{j}_x({\bf r}_1)(1-\Pi) \hat{j}_x({\bf r}_2)|G\rangle \\
   &={\rm Tr}\ \left[\Pi\hat{j}_x({\bf r}_1)(1-\Pi) \hat{j}_x({\bf r}_2)\right],\\
&= \sum_{{\bf r}_3 {\bf r}_4}P_{{\bf r}_2 {\bf r}_4}P_{{\bf r}_4{\bf r}_1}P_{{\bf r}_1{\bf r}_3}P_{{\bf r}_3{\bf r}_2}\left(x_1-x_4\right)\left(x_2-x_3\right) \\
& \quad - \sum_{{\bf r}_4}P_{{\bf r}_2 {\bf r}_4}P_{{\bf r}_4{\bf r}_1}P_{{\bf r}_1{\bf r}_2}\left(x_1-x_4\right)\left(x_2-x_1\right) \,,
\end{aligned}
\right.
\label{eq:j-j}
\end{equation}
where $P_{{\bf r}'{\bf r}}\equiv \langle G|c^\dagger_{{\bf r}'} c_{\bf r}|G\rangle $ is the two-point function and $x_i$ is the $x$ coordinate of position ${\bf r}_i$. Here, we used the definition of the current density operator for the third equality~\footnote{For a Bogoliubov deGenne Hamiltonian, $H'$ takes the fermion bi-linear form after relabelling a set of fermion annihilation operators as creation operators and vice versa.}.
(See Appendix for further details.)
Hence for the approximate Hamiltonian ${H}'$, the current-current correlation function at zero temperature consists of an appropriately weighted combination of quadrilateral loops and triangular loops of two-point functions.

Note that $L^\triangle_{jkl}$ and  $L^\Box_{ijkl}$ defined in Eqs.~\eqref{eq:triangle} and \eqref{eq:quad} involve samples of the Green's functions $\widetilde{P}_{jk}|_{\alpha}$ typically coming from a determinant quantum Monte Carlo (DQMC) calculation.
By processing the loop operators during the sampling process and avoiding an a posteriori Monte Carlo averaging, we quickly pass these fluctuation-laden data, which encodes (partial) information of the current-current correlation function, to the machine learning step, see Fig.~\ref{fig:network_architecture}(a).
Clearly, the loop operators $L^\Box_{jklm}$ and $ L^\triangle_{jkl}$ built from individual Monte Carlo samples and only for short-ranged loops, cannot replace a rigorous calculation of the current-current correlation function, especially for a gapless system. But we anticipate the QLT consisting of the triangular and quadrilateral loops to serve as a proxy for the current-current correlation function containing qualitative information regarding longitudinal transport directly in the imaginary time data. Such a proxy is particularly desirable since traditional approaches,  based on an explicit construction of time-displaced Green's functions, are costly and frequently require numerical stabilization \cite{Hirsch1988, Santos2003}, whereas QLT only demands equal-time correlations, readily available in DQMC simulations.


\noindent{\em Models and Results.-- }
To test the potential of the QLT+QMC approach for efficiently detecting qualitative differences in the transport from equal-time Green's function data, we consider two paradigmatic model systems that host superconductivity in parts of their respective phase diagrams -- the attractive Hubbard model and a spin-fermion model of a quantum critical metal.
Since both models are two-dimensional lattice models, we note that there are two subtleties to detecting superconductivity  in two spatial dimensions (2D). First, the superconducting order parameter is not readily accessible in a QMC simulation and one has to follow a carefully defined limiting process to obtain the superfluid density from current-current correlation functions \cite{Scalapino1993, Scalapino1992}. Second, the superconducting phase transition in 2D is of Kosterlitz-Thouless (KT) type and hence the superconducting transition is signaled by the superfluid density exceeding the  critical KT value \cite{KT1973}. Prior to this transition, the onset of superconducting fluctuations and a regime of diamagnetism is indicated by a sign change in the orbital magnetic response \cite{Schattner2016}. Given the explicit tie between the QLT and the zero-frequency current-current correlation functions discussed earlier for the simple gapped Hamiltonian, one can readily anticipate that the QLT approach will provide enough information on superconducting fluctuations so that the artificial neural network (ANN) fed with this data will be able to recognize the onset of the diamagnetic regime that precedes  superconductivity.

We start by considering what is probably considered the simplest model for superconductivity -- the negative-$U$ Hubbard model on the square lattice \cite{Scalettar1989, Scalapino1992}
\begin{eqnarray}
H = &-&\underset{\left\langle ij \right\rangle, s}{\sum} \left( c^\dagger_{j,s}c_{i,s} +  c^\dagger_{i,s}c_{j,s} \right) -  \mu\underset{i}{\sum} \left(n_{i,\uparrow} +n_{i,\downarrow}\right) \,, \nonumber\\
&+& U \underset{i}{\sum} \left(n_{i,\uparrow}-\frac{1}{2}\right) \left(n_{i,\downarrow}-\frac{1}{2}\right)
\label{eq:hubbard}
\end{eqnarray}
where $c^\dagger_{i,s}$ is an electron creation operator at site $i$ with spin $s =\, \uparrow, \downarrow$  and $n_{i,s}=c^\dagger_{i,s}c^{\phantom\dagger}_{i,s}$ is the electron density operator. $U=-\left|U\right|<0$ is the attractive interaction strength and $\mu$ is the chemical potential.  Without loss of generality, we set $U=-8$ and tune the electron density $\left\langle n\right\rangle = \left\langle n_\uparrow\right\rangle+ \left\langle n_\downarrow\right\rangle  \simeq 0.9$ slightly below half filling.
Numerically, we study a system of $8\times 8$ sites. For each site, we  build and collect site-touching quantum loops $L^\Box$ and $ L^\triangle$. The so constructed QLT data forms a field of principle input vectors $x$ for a shallow ANN, see Fig.~\ref{fig:network_architecture}(a).

\begin{figure}[t]
\includegraphics[scale=.43]{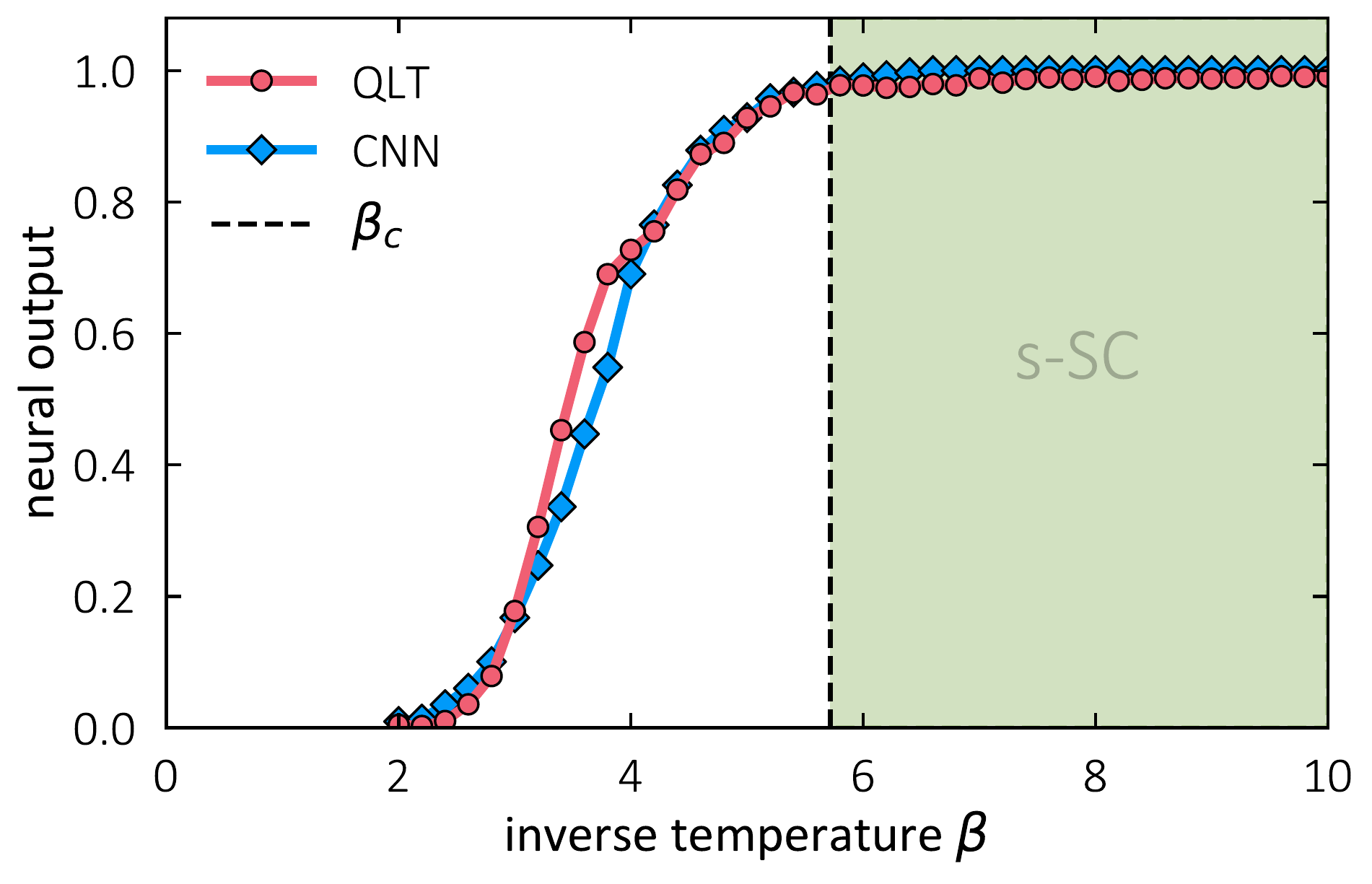}
\caption{
{\bf Negative-U Hubbard model.} The neural outputs of the quantum loop topography (QLT) and convolutional neural network (CNN) architectures (Fig.~\ref{fig:network_architecture}) for superconducting transport versus the inverse temperature $\beta$. While the inputs for the deep CNN are the unprocessed Green's functions $P(\mathbf{r},\mathbf{r}')$ from the square-lattice negative-$U$ Hubbard model in Eq.~\eqref{eq:hubbard},
we preprocess the input for the shallow ANN of the QLT architecture in the form of the quantum loops in Eq.~\eqref{eq:triangle} and \eqref{eq:quad}. Both architectures are trained with samples at low temperature $\beta=20$ representing superconducting transport and high temperature $\beta=2$ representing normal state transport. Then the resulting architectures are applied towards the interpolating temperatures. The vertical dashed line indicates $\beta_c = 1/T_c \approx 5.8$ \cite{Scalettar1989, Denteneer1993} and defines the s-wave superconductor phase (green shaded region).}\label{fig:hubbard}
\end{figure}

In an initial training step, the ANN is optimized with a training set consisting of about 20,000 samples obtained from the superconducting phase at low temperature ($\beta=20$) and the normal phase at high temperature ($\beta=2$).
For training, we employ a cross-entropy cost function and L2 regularization to avoid over-training and a mini-batch size of 10. We also reserve an independent validation set of $10\sim 20\%$ of the training data set for validation purposes (such as learning speed control and termination \cite{MLbook}).
After this training step, the QLT input from a range of $\beta$ interpolating between the two training points is classified using the optimized ANN, with the result summarized in Fig.~\ref{fig:hubbard}.
Clearly, the ANN transport classification of the two phases is achieved with a high confidence  $>99\%$ in the low and high temperature limits. In between, it indicates a smooth onset around the temperature $T_{\rm fluct}=\beta^{-1} \sim 0.28$, which we interpret as the onset of superconducting fluctuations and therefore expect it to be slightly higher than the reported critical temperature for the KT transition $T_c\simeq 0.17$ \cite{Scalettar1989, Denteneer1993, Scalapino1991}. As we argued above, the ANN has no means of determining the exact KT transition temperature, and this is the best performance we can in fact anticipate.

\begin{figure}
\includegraphics[scale=.43]{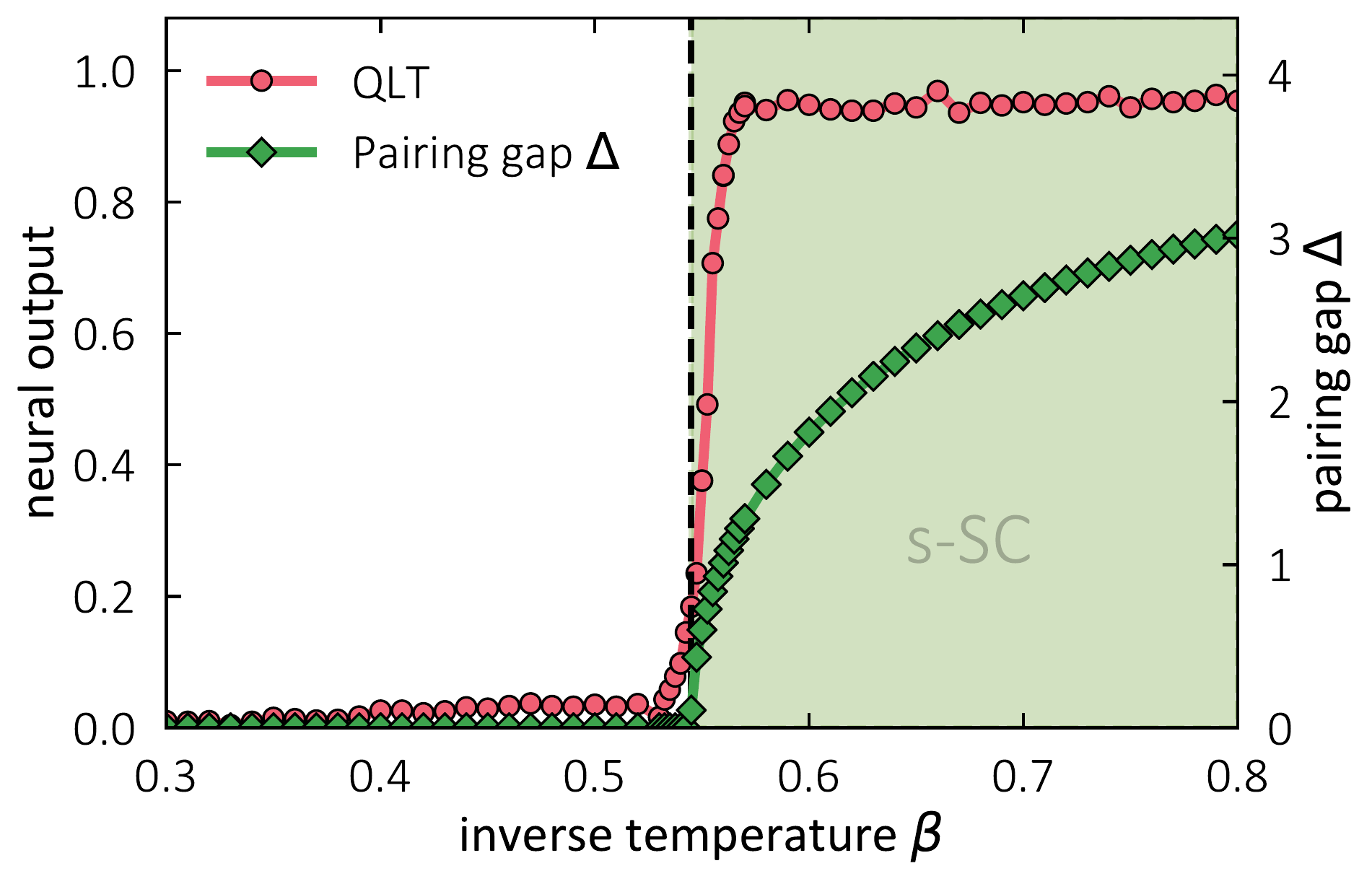}
\caption{{\bf Mean-field transition of the negative-U Hubbard model.}
		Comparison of the mean-field $s$-wave pairing gap $\Delta$ (red dots)
		and the neural output from the QLT+QMC approach.
		The consistent onset of both functions demonstrates that the machine learning approach is indeed sensitive
		to the onset of superconductivity.
		The vertical dashed line indicates the mean-field transition temperature $\beta_c\sim 0.545$.}
		\label{fig:mlmft}
\end{figure}

We benchmark our QLT approach against two alternative approaches.
First, we consider the mean-field theory of the negative-$U$ Hubbard model and compare our QLT+QMC approach for this
mean-field model against exact analytical results. Since the mean-field theory cannot capture fluctuations, the change of transport properties and the superconducting transition strictly coincide within the mean-field theory. Hence, we anticipate the assessment of superconductivity within the QLT approach to coincide with the mean-field transition. To test this, we solve the self-consistent gap equations at  each inverse temperature $\beta$ to obtain the $s$-wave pairing gap $\Delta(\beta)$. We then sample via finite-temperature Monte Carlo simulations, using the corresponding BdG mean-field model with the respective $\Delta(\beta)$, and generate a Markov sequence of the quasi-particle occupation number, which is then fed into the QLT approach.
 The resulting phase diagram (for $U=-8$ and $\mu=-0.5$) is  shown in Fig.~\ref{fig:mlmft} where we used $\beta=0.3$ and $\beta=0.8$ as the high and low temperature training points, respectively.
Indeed, the QLT assessment of superconductivity shows a sharp onset at the mean-field superconducting transition.

Our second benchmark is to compare against an alternative numerical approach where the entire, unprocessed Green's functions
$\widetilde{P}_{jk}|_{\alpha}$ for all $j,k$'s are used as input for a convolutional neural network, Fig.~\ref{fig:network_architecture}(b), akin to previous work~\cite{Simon2016} that demonstrated the feasibility of such an approach to  locate symmetry-breaking phase transitions in quantum many-fermion systems. Note that these data sets are 4-dimensional, $L^2 \times L^2$, and hence considerably larger than the condensed quasi-two-dimensional QLT loop vector fields, $L^2 \times D(d_c)$, where $D(d_c)$ denotes the dimension of a loop vector for given maximal loop length $d_c$.
Fig.~\ref{fig:hubbard} shows the direct comparison of the QLT+shallow ANN approach versus such a CNN setting. With the two techniques giving essentially the same result, we conclude that  both approaches are indeed capable of detecting the onset of superconducting fluctuations from raw data of equal-time Green's functions. The QLT approach, however, succeeds in doing so with a significantly smaller data set, which is of enormous practical advantage.

\begin{figure}[b]
\includegraphics[scale=.43]{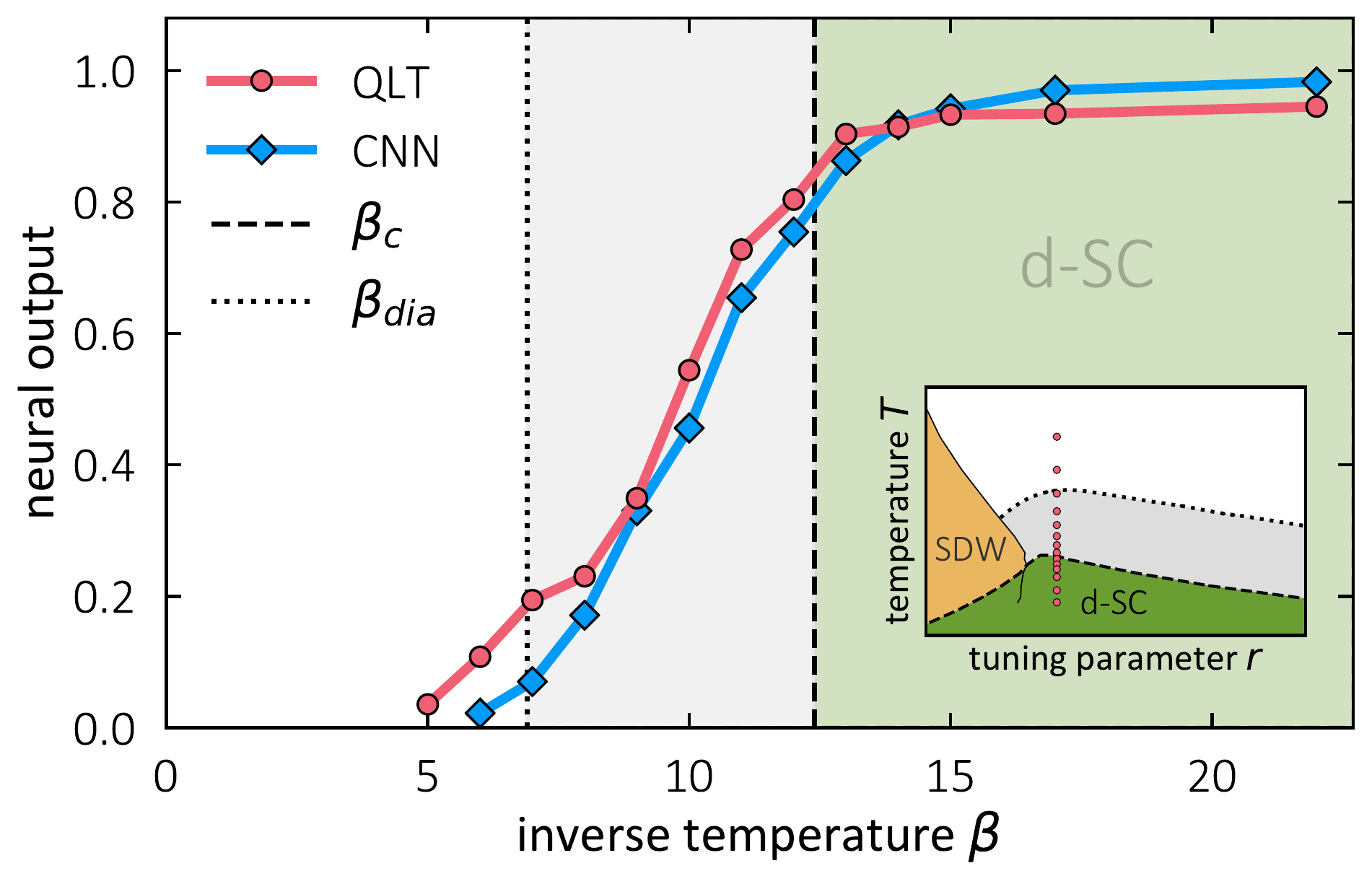}
\caption{Neural network output for the superconducting transition near a {\bf metallic quantum critical point}. The training points were at $\beta=30$ for superfluid transport and $\beta=5$ for metallic transport.
The vertical lines indicate the superconducting transition temperature $\beta_c\sim 12.5$ (dashed) derived from the superfluid density measurements and the onset of diamagnetic fluctuations $\beta_{\textrm{dia}} \sim 6.9$ (dotted) where the orbital magnetic susceptibility changes sign \cite{Schattner2016}. The inset (modified from \cite{Schattner2016}) illustrates the chosen finite-temperature scan cutting into the superconducting dome.} 
\label{fig:afmetal}
\end{figure}
As a second principle example, we now turn to a model in which superconductivity arises from the quantum critical fluctuations in the vicinity of a spin-density wave (SDW) transition \cite{Berg2012,Schattner2016,Gerlach2017,Berg2018}. Specifically,
we consider a square lattice spin-fermion model $S = S_\psi + S_\varphi + S_\lambda$ \cite{Schattner2016}, in which two flavors of spin$-1/2$ fermions are coupled to an easy-plane SDW order parameter $\vec{\varphi}$ at wave vector $\mathbf{Q}=(\pi, \pi)$,
\begin{eqnarray}
S_\psi &=&  - \int_{\tau, \mathbf{r}, \mathbf{r'}} \sum_{s, \alpha} \left[ \left(\partial_\tau - \mu\right)\delta_{\mathbf{r}\mathbf{r'}} - t_{\alpha \mathbf{r}\mathbf{r'}} \right] \psi_{\alpha \mathbf{r}s}^\dagger \psi_{\alpha \mathbf{r'}s} \nonumber\\
S_\lambda &=& \lambda \int_{\tau, \mathbf{r}} e^{i \mathbf{Q}\cdot \mathbf{r_i}} \vec{\varphi}_\mathbf{r} \cdot \left( \psi_{a\mathbf{r}s}^\dagger \vec{\sigma}_{ss'}  \psi_{b\mathbf{r}s'} + \textrm{h.c.} \right) \nonumber\\
S_\varphi &=& \int_{\tau,\mathbf{r}} \frac{1}{2c^2} \left(\partial_\tau \vec{\varphi}\right)^2 + \left(\nabla \vec{\varphi} \right)^2 + \frac{r}{2}\vec \varphi^2 + \frac{u}{4} (\vec \varphi^2)^2,
\label{eq:afmetal}
\end{eqnarray}
where $\alpha=a,b$ labels the two fermion flavors, and $s = \uparrow, \downarrow$ denotes the spin. The nearest-neighbor fermion hopping amplitudes are chosen as $t_{a,x}=t_{b,y}=1$ and $t_{a,y}=t_{b,x}=0.5$. We further set the Yukawa coupling $\lambda=3$, chemical potential $\mu=0.5$, $u=1$ and the bare bosonic velocity to be $c=2$.
Studying a system of size $8\times 8$, we tune the dynamics of the SDW order parameter to the vicinity of the SDW transition, $r=10.35$, around which a finite-temperature superconducting dome is found \cite{Schattner2016}.

The results of a finite-temperature scan cutting into the superconducting dome are shown in Fig.~\ref{fig:afmetal}. It clearly confirms that the QLT+QMC approach can indeed correctly detect the onset of superconducting fluctuations in this spin-fermion model. This is evident by the fact that the neural network output is turning on where earlier studies of current-current correlations \cite{Schattner2016,Lederer2017} as a proxy for superconductivity found the onset of diamagnetic fluctuations (dotted line) \cite{Schattner2016}.
We also find very good agreement with the alternative numerical approach of feeding the unprocessed Green's function data into a CNN -- albeit using only the dimensionally reduced QLT data.
Further note that the QLT+QMC approach achieved the above detection of superconducting transport using only $O(10)$ uncorrelated samples of the QMC Markov chain \footnote{This includes the configurations needed in constructing the quantum loops as well as for streamlining the accuracy.}, a huge reduction over the number of samples used in the traditional superfluid density calculation \cite{Hong2016, Schattner2016}.


\noindent{\em Conclusions.-- }
To summarize, we have introduced a feature selection protocol for machine learning longitudinal transport and demonstrated that such a QLT preprocessing step allows to identify the onset of superconductivity in quantum many-fermion systems at considerably lower numerical cost than traditional approaches.
Comparing to other machine learning approaches, we showed that the response theory guided QLT+QMC approach performs just as well as the much more involved CNN approach rooted in image recognition techniques while using only a fraction of the Monte Carlo data. The major advantage of the QLT+QMC approach is that it is motivated by quantum statistical physics considerations and therefore allows for a much better intuitive understanding of its capabilities. A further advantage is that the  QLT is semi-locally defined and as such it does not require e.g. translational symmetry.
In an explicit application, we showed that the QLT+QMC approach can detect correctly transport signatures of superconducting fluctuations by comparing the neural network outcome to rigorous traditional measurements of the superfluid density.
Looking ahead, we anticipate that QLT+QMC will be even more valuable in detecting states without traditional representation that are nevertheless defined through non-trivial transport properties such as non-Fermi liquids.


{\it Acknowledgements.--}
We acknowledge useful discussions with Erez Berg, Yi-Ting Hsu, and Hong Yao. YZ and E-AK acknowledge the support from the U.S. Department of Energy, Office of Basic Energy Sciences, Division of Materials Science and Engineering under Award DE-SC0018946.
The Cologne group acknowledges partial support from the Deutsche Forschungsgemeinschaft (DFG, German Research Foundation) -- Projektnummer 277101999 -- TRR 183 (project B01).
The numerical simulations were performed on the JUWELS cluster at FZ J\"ulich and the CHEOPS cluster at RRZK Cologne.

\bibliography{refs,t+X-NSF}
\appendix

\section{Current-current correlations and quantum loop topography}

In this appendix, we explore the connection between the zero-frequency current-current correlation in Eq.~\eqref{eq:Lambda}
\begin{equation}
\left. \begin{aligned}
\Lambda&\left({\bf r}_1, {\bf r}_2;\omega=0\right) = \int d\tau \left\langle \hat{j}_x({\bf r}_1,\tau)\hat{j}_x({\bf r}_2,0) \right\rangle  \\
&=\sum_{n\ne G} \frac{\left\langle G \left| \hat{j}_x({\bf r}_1)  \left|n\left\rangle  \right\langle n\right| \hat{j}_x({\bf r}_2)\right| G\right\rangle}{E_n-E_G}
\end{aligned}
\right.
\end{equation}
and the products of two-point correlators trailing loops, especially in the presence of superconductivity. For simplicity, we focus on a mean-field Hamiltonian with a flat band $H'=1/2-\hat P$, $\hat P=\sum_{{\bf r}'{\bf r}}P_{{\bf r}'{\bf r}}a^\dagger_{{\bf r}'} a_{\bf r}=\sum_{\mu}a^\dagger_\mu a_\mu$. Then,
\begin{equation}
\left. \begin{aligned}
\Lambda\left({\bf r}_1, {\bf r}_2;\omega=0\right)&
=\sum_{\mu, \nu}\left\langle \mu \left| \hat{j}_x({\bf r}_1)  \left|\nu\left\rangle  \right\langle \nu\right|\hat{j}_x({\bf r}_2)\right| \mu\right\rangle \\
= \mbox{tr}&\left[\Pi \hat{j}_x({\bf r}_1) \left(1-\Pi\right) \hat{j}_x({\bf r}_2)\right]
\end{aligned}
\right.
\label{eq:app1}
\end{equation}
where $\left|\mu\right\rangle$ ($\left|\nu\right\rangle$) are the single-particle states that are occupied (empty) in the ground state $\left|G\right\rangle$ of $H'$. It follows that $\epsilon_\mu=-1/2$, $\epsilon_\nu=1/2$, and $1-\hat P=\sum_{\nu}a^\dagger_\nu a_\nu=\sum_{\nu}\left|\nu\right\rangle\left\langle\nu\right|$. We note that the arguments also apply for a mean-field Bogoliubov deGenne Hamiltonian of a superconductor
\begin{equation}
\left. \begin{aligned}
H_{BdG} & =  \sum_{\bf{r}'\bf{r}}t_{\bf{r}'\bf{r}}c_{\bf{r}'}^{\dagger}c_{\bf{r}}+\Delta_{\bf{r}'\bf{r}}c_{\bf{r}'}^{\dagger}c_{\bf{r}}^{\dagger}+\mbox{h.c.}\\
=\frac{1}{2}\sum_{\bf{r}'\bf{r}}&t_{\bf{r}'\bf{r}}\left(a_{\bf{r}'}^{\dagger}a_{\bf{r}}-\tilde{a}_{\bf{r}}^{\dagger}\tilde{a}_{\bf{r}'}\right)
+\Delta_{\bf{r}'\bf{r}}\left(a_{\bf{r}'}^{\dagger}\tilde{a}_{\bf{r}}-a_{\bf{r}}^{\dagger}\tilde{a}_{\bf{r}'}\right)+\mbox{h.c.}
\end{aligned},
\right.
\end{equation}
where $a_{\bf{r}}^\dagger=\tilde{a}_{\bf{r}}=c_{\bf{r}}^\dagger$ is the creation operator of the original electrons. With the new basis of both the electron and hole operators $a$ and $\tilde{a}$, the Hamiltonian $H_{BdG}$ again takes a fermion bi-linear form. In the following, we will drop the tilde and denote $a(\bf{r})$ and $\tilde{a}(\bf{r})$ with their respective electric charge $q=\pm 1$ that we contain in the label $\bf{r}$.

Commonly, the $\vec{x}$-direction current operator at position ${\bf r}$ is
\[
	\hat{j}_x({\bf r})=-i\left[H'({\bf r}),\hat x\right]=\sum_{{\bf r}'}iP_{{\bf r}'{\bf r}}c_{{\bf r}'}^\dagger c_{{\bf r}}\left(x-x'\right)+\mbox{h.c.} \,,
\]
where $x$ is the $\vec{x}$-direction component of ${\bf r}$. In the case of Bogoliubov deGenne Hamiltonian, however, we need to incorporate the sign of the charge and generalize the current operator as ${\bf r}$ is
\[
	\hat{j}_x({\bf r})=-\frac{i}{2}\left(\hat{q}\left[H'({\bf r}),\hat x\right]+\left[H'({\bf r}),\hat x\right]\hat{q}\right) \,.
\]
Putting this into Eq.~\eqref{eq:app1}, we have
\begin{equation}
\left.
\begin{aligned}
\mbox{tr}&\left[\Pi \hat{j}_x({\bf r}_1) \left(1-\Pi\right) \hat{j}_x({\bf r}_2)\right]  \\
=& \frac{1}{4} \sum_{{\bf r}_3 {\bf r}_4 {\bf r}'_1 {\bf r}'_2{\bf r}'_3{\bf r}'_4}P_{{\bf r}'_4 {\bf r}_4}P_{{\bf r}'_1{\bf r}_1}\left(P_{{\bf r}'_3{\bf r}_3}-\delta_{{\bf r}'_3{\bf r}_3}\right)P_{{\bf r}'_2{\bf r}_2} \\&\times\left(x_1-x'_1\right)\left(x_2-x'_2\right)\left(q_1+q'_1\right)\left(q_2+q'_2\right)
\\&\times\left\langle0\left|a_{{\bf r}_4}a^\dagger_{{\bf r}'_1} a_{{\bf r}_1} a^\dagger_{{\bf r}'_3} a_{{\bf r}_3} a^\dagger_{{\bf r}'_2} a_{{\bf r}_2} a^\dagger_{{\bf r}'_4}\right|0\right\rangle \\
=& \sum_{{\bf r}_3 {\bf r}_4}P_{{\bf r}_2 {\bf r}_4}P_{{\bf r}_4{\bf r}_1}P_{{\bf r}_1{\bf r}_3}P_{{\bf r}_3{\bf r}_2}\\
&\times\left(x_1-x_4\right)\left(x_2-x_3\right)\left(q_1+q_4\right)\left(q_2+q_3\right)/4 \\
 &- \sum_{{\bf r}_4}P_{{\bf r}_2 {\bf r}_4}P_{{\bf r}_4{\bf r}_1}P_{{\bf r}_1{\bf r}_2}\\
 &\times\left(x_1-x_4\right)\left(x_2-x_1\right)\left(q_1+q_4\right)\left(q_2+q_1\right)/4
\label{eq:app2}\end{aligned}\right.
\end{equation}

On the other hand, the expectation value of the two-point correlator is given by
\begin{equation}
\left. \begin{aligned}
\left\langle a_{{\bf r}'}^{\dagger}a_{{\bf r}}\right\rangle  &= \sum_{\mu}\left\langle \mu|a_{{\bf r}'}^{\dagger}a_{{\bf r}}|\mu\right\rangle \\
  =\sum_{\tilde{{\bf r}}' \tilde{{\bf r}}}&P_{\tilde{{\bf r}}\tilde{{\bf r}}'}\left\langle 0|a_{\tilde{{\bf r}}'}a_{{\bf r}'}^{\dagger}a_{{\bf r}}a_{\tilde{{\bf r}}}^{\dagger}|0\right\rangle  =P_{{\bf r}{\bf r}'}
\end{aligned}\right.
\label{eq:j-jbdg}
\end{equation}
Therefore, Eq.~\eqref{eq:app2} can be expressed in terms of products of two-point correlators trailing triangles and quadrilaterals. In particular, in terms of the original electron correlators $P_{\bf{r}\bf{r}'}=\left\langle c^\dagger_{\bf{r}}c_{\bf{r}'} \right\rangle$, $q_{\bf{r}}=1$ for all and Eq.~\eqref{eq:j-jbdg} returns to Eq.~\eqref{eq:j-j}.

\end{document}